\documentclass{llncs}
\usepackage{graphics,amsfonts}
\usepackage[latin1]{inputenc}
\usepackage{graphicx}
\usepackage{appendix}
\usepackage{amsfonts}
\usepackage{amsmath,amscd}
\usepackage{amssymb}
\usepackage{makeidx}
\usepackage{algpseudocode}
\usepackage{algorithm}
\begin{document}
\title{A New Guess-and-Determine Attack on the \emph {A5/1} Stream Cipher}
\author{Jay Shah \and Ayan Mahalanobis}
\institute{Indian Institute of Science Education and Research, Pune, India\\
\email{jayshah\_89@hotmail.com; ayan.mahalanobis@gmail.com}}
\maketitle
\begin{abstract}
In Europe and North America, the most widely used stream cipher to ensure privacy and confidentiality of conversations in GSM mobile phones is the \emph {A5/1}. In this paper, we present a new attack on the \emph {A5/1} stream cipher with an average time complexity of $2^{48.5}$, which is much less than the brute-force attack with a complexity of $2^{64}$. The attack has a 100\% success rate and requires about 5.65GB storage. We provide a detailed description of our new attack along with its implementation and results.
\newline
\textbf{Keywords:} \emph{A5/1}, GSM, guess-and-determine attack, stream ciphers
\end{abstract}

\section{Introduction}
The most widely used stream cipher to ensure privacy and confidentiality of conversations in GSM mobile phones in Europe and North America is the \emph {A5/1}. The \emph {A5/1} was developed in 1987, then GSM was not  considered for use outside Europe. The description of the \emph {A5/1} was initially kept secret, but its design was disclosed in 1999 by reverse engineering \cite{[Bri]}. The GSM organization later confirmed the correctness of the algorithm \cite{[Bir]}.

There are multiple versions of the encryption algorithm which belong to the \emph {A5} family: \emph {A5/0} is a dummy cipher with no encryption; \emph {A5/1} is the original \emph {A5} algorithm ensures over-the-air communication privacy and confidentiality of conversations in GSM mobile phones; \emph {A5/2} is an intentionally weaker encryption algorithm created for export; while \emph {A5/3} is a strong encryption algorithm created as part of the 3rd Generation Partnership Project \emph{(3GPP)} which is currently responsible for maintaining and developing GSM technical specifications around the world \cite{[Hill]}.

Anderson \cite{[And]}, Golic \cite{[Golic]} and Babbage \cite{[Bab]} were the pioneers in cryptanalyzing the \emph {A5/1} encryption algorithm when only a rough outline of the \emph {A5/1} was leaked. After \emph {A5/1} was reverse engineered, it was analyzed by Biryukov, Shamir and Wagner \cite{[Bir]}; Biham and Dunkelman \cite{[BD]}; Ekdahl and Johansson \cite{[Ek]}; Maximov, Johansson and Babbage \cite{[Max]}; Barkan and Biham \cite{[Bar]}; Keller and Seitz \cite{[KS01]}; and a few other researchers. 

\subsection{Current Research}
Several attacks on the \emph {A5/1} stream cipher were designed in the past twenty years, but only a few of those were implemented. Attacks on the GSM protocol can work even if the network supports only \emph {A5/1} or \emph {A5/3} encryption, as long as the mobile phone supports \emph {A5/2} encryption. The main flaw that allows the implementation of these attacks is that the same key is used regardless of whether the phone encrypts using \emph {A5/1}, \emph {A5/2}, or \emph {A5/3} algorithm. Therefore, the attacker can mount a man-in-the-middle attack, in which the attacker impersonates the mobile to the network, and the network to the mobile (by using a fake base station). The attacker might use \emph {A5/1} for communication with the network and \emph {A5/2} for communication with the mobile. But due to the flaw, both algorithms encrypt using the same key. The attacker can obtain the key through a passive attack on \emph {A5/2}. The attacker who is in the middle can eavesdrop, change the conversation, perform call theft, etc. The attack applies to all traffic including short message service (SMS) \cite{[Bar]}.

\subsection{Our Contributions}
In this paper we describe a new \emph{guess-and-determine} attack on the \emph {A5/1} stream cipher. This attack has an average time complexity of $2^{48.5}$, which is much less than a brute-force attack of $2^{64}$. For every \emph{possible} 19 bits of the register $R_1$, we require storage of 100MB to determine registers $R_2$ and $R_3$ given a known keystream KS. Our attack can be briefly described as follows: we assume that the register $R_1$ is full with 19 bits and registers $R_2$ and $R_3$ will be filled progressively as the attack progresses. At any stage of this attack, $R_1$ is completely filled and $R_2$ and $R_3$ are partially filled. We call these states as \textbf{state candidates}. Once all three registers are completely filled, we call that state candidate a \textbf{complete state candidate}. 
This attack has a 100\% success rate and requires about 5.65GB storage. With the knowledge of only 11 bits of the known keystream, the attack algorithm is able to determine a set of 64-bit \emph{complete state candidates} which may contain the key. With every additional clocking round of the attack, the number of complete state candidates increases. Thus, the probability of finding the key among all the complete state candidates increases with every additional round after 11 clocking rounds. We provide a detailed description of our new attack along with its implementation and results in Sections 4, 5 and 6.

\section{Description of the \emph {A5/1} Stream Cipher}
\begin{figure}[!htb]
\begin{center}
\hspace*{-1cm}
\includegraphics[trim=0cm 9cm 0cm 7cm, clip=true, scale=0.42]{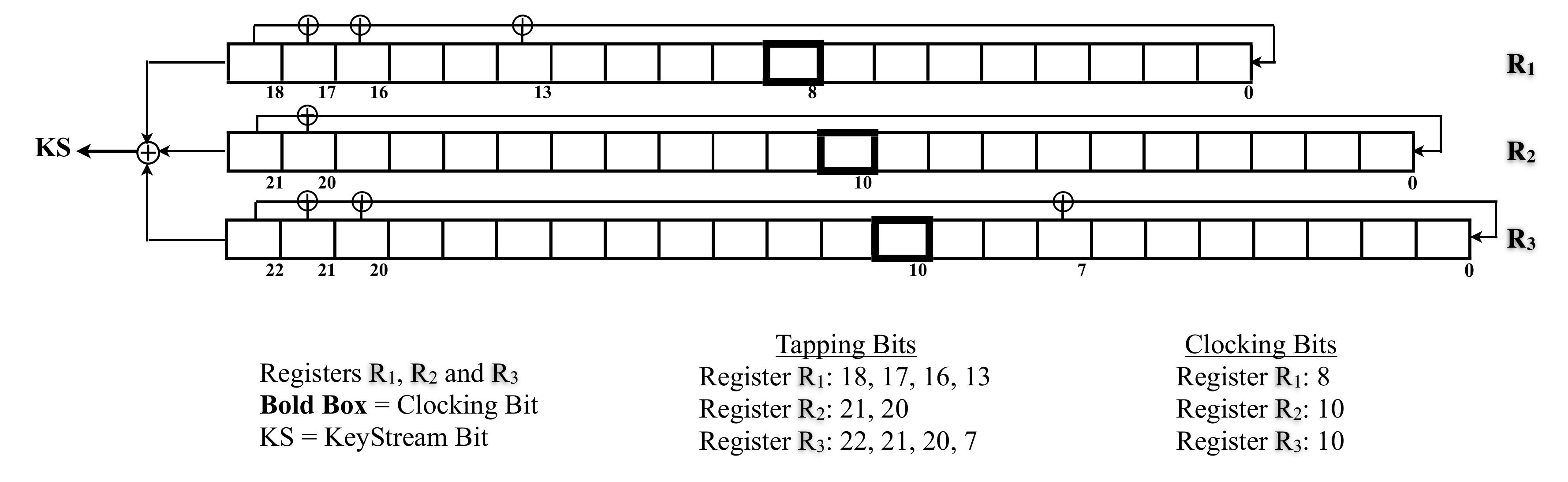}
\caption{\emph{A5/1} Stream Cipher}
\label{a}
\end{center}
\end{figure}
The \emph {A5/1} stream cipher is built from three short linear feedback
shift registers (LFSRs) of lengths $19$, $22$, and $23$ bits, denoted by $R_{1}$, $R_{2}$ and $R_{3}$ respectively. The rightmost bit in each register is labeled as bit zero. The tapping bits of $R_{1}$ are at bit positions $13$, $16$, $17$, $18$; the tapping bits of $ R_{2} $ are at bit positions $20$, $21$; and the tapping bits of $R_{3}$ are at bit positions $7$, $20$, $21$, $22$ (Table \ref{par}). 

\begin{table}[ht]
\caption{The \emph{A5/1} Register Parameters} 
\centering      
\begin{tabular}{c c c c c}  
\hline\hline                        
Register & Register Length & Clocking Bit & Primitive Polynomials & Tapping Bits \\ [0.5ex] 
\hline                    
$R_{1}$ & 19 bits & 8 & $1 + x + x^2 + x^5 + x^{19}$ & 18, 17, 16, 13 \\ 
$R_{2}$ & 22 bits & 10 & $1 + x + x^{22}$ & 21, 20 \\ 
$R_{3}$ & 23 bits & 10 & $1 + x + x^2 + x^{15} + x^{23}$ & 22, 21, 20, 7  \\ [1ex]       
\hline     
\end{tabular} 
\label{par}  
\end{table}

The tapping bits are predetermined according to the corresponding primitive polynomials for the registers. A polynomial of degree $n$ over the finite field $GF(2)$ (i.e., with coefficients either $0$ or $1$) is primitive if it has polynomial order $2^{n} -1$.
For each register, when the register is clocked, its tapping bits are XORed together and the result is stored in the rightmost bit of the left-shifted register. The three registers are maximal length LFSRs with periods $2^{19} - 1$, $2^{22} - 1$, and $2^{23} - 1$ respectively \cite{[period]}.

The \emph {A5/1} keystream generator works as follows \cite{[Bri]}. First, an initialization phase is run. At the start of this phase, all bits of the registers are set to $0$. Then the key setup and the \emph{Initialization Vector (IV)} setup are performed. During the initialization phase, all three registers are clocked and the key bits followed by the \emph{IV} bits are XORed with the \emph{most significant bits} (MSBs) of all three registers. Thus, the initialization phase takes an overall of $64 + 22 = 86$ clock-cycles after which initial state $\emph S_{i}$ is achieved.

Based on this initial state $\emph S_{i}$, a warm-up phase is performed where the generator is clocked for $100$ clock-cycles and the output is discarded. This results directly in state $\emph S_{w}$ producing the first output bit $101$ clock-cycles after the initialization phase. During the warm-up phase and the stream generation phase, the registers $R_{1}$, $R_{2}$, and $R_{3}$ are clocked irregularly according to the majority function rule \cite{[Donald]} depending on the \emph{clocking bits} (CBs) of the registers. The majority function is a function from $n$ inputs to one output. The value of the operation is true when at least $\frac{n}{2}$ arguments are true, and false otherwise.

The registers are clocked in a stop/go fashion using the following majority rule: Each register has a single clocking bit (bit $8$ for $R_{1}$, bit 10 for $R_{2}$, and bit $10$ for $R_{3}$) which decides the clocking pattern for its respective register. In each clock cycle, the majority function of the clocking taps is calculated and only those registers whose CBs agree with the majority function are clocked. At each step either two or three registers are clocked, and that each register has a probability of moving $3$ out of $4$ times. It is this clocking pattern which makes the stream cipher generate output bits which are random.

During encryption, a total of four cases are possible for clocking pattern of the registers. They are:
\begin{center}
Case 1: $ CB_1 = CB_2 \neq CB_3$ (Clock $R_{1}$ and $R_{2}$ only)
\newline Case 2: $CB_1 \neq CB_2 = CB_3$ (Clock $R_{2}$ and $R_{3}$ only)
\newline Case 3: $CB_1 = CB_3 \neq CB_2$ (Clock $R_{1}$ and $R_{3}$ only)
\newline Case 4: $CB_1 = CB_2 = CB_3$ (Clock  all three registers)
\newline
\end{center}
where $CB_i$ denotes the \emph{clocking bit} for register $i$; $i = (1, 2, 3)$.
\newline\newline
After clocking, an output bit is generated from the values of $R_{1}$, $R_{2}$, and $R_{3}$ by XORing their \emph{most significant bits} (MSBs), as shown in Equation \ref{eq}. This XORed bit is called the \emph {keystream bit} (KS).
\begin{equation}
R_{1}[18] \oplus\; R_{2}[21] \oplus R_{3}[22] = KS[i],
\label{eq}
\end{equation}
where KS[i] denotes the $i\textsuperscript{th}$ keystream bit, $i=0$ on initialization and increases by 1 after every clocking round.

After warm-up phase, the \emph{A5/1} produces 228 output bits. For every clock cycle, 114 bits are used to encrypt uplink traffic, while the remaining 114 bits are used to decrypt downlink traffic \cite{[Gend]}.

\section{Known Attacks on the \emph{A5/1}}
Here we discuss the known \emph{guess-and-determine} attacks on the \emph{A5/1}. A guess-and-determine attack \cite{[Men]} is a known-plaintext attack on stream ciphers where the attacker guesses some bits of the cipher and the remaining bits are determined from the known keystream bits. The known plaintext attack is an attack model where the attacker has access to both the plaintext and its encrypted ciphertext. This can be used to reveal the secret key used for encrypting the known plaintext to the known ciphertext. These include Anderson's Attack \cite{[And]}, Golic's Attack \cite{[Golic]}, Biham-Dunkelman's Attack \cite{[BD]}, Keller-Seitz's Attack \cite{[KS01]} and Gendrullis-Novotny-Rupp's Attack (also known as the Modified Keller-Seitz Attack) \cite{[Rupp]}. All these attacks assume to have 64 bits of the keystream (KS) known.

\subsection{Guess-and-Determine Attacks}
The first guess-and-determine attack on the \emph{A5/1} was proposed by Anderson \cite{[And]}. Anderson suggested to guess all bits of registers $R_{1}$ and $R_{2}$ and the lower half of register $R_{3}$ (i.e., 19 + 22 + 11 = 52 bits), to determine the remaining bits of $R_3$ by Equation \ref {eq}. In the worst-case, each of the $2^{52}$ determined state candidates need to be verified against the known keystream. This attack was not implemented as Biham-Dunkelman's Attack and Keller-Seitz's Attack had lesser complexity.

Golic proposed an attack that has a complexity of $2^{40}$ linear equations sets \cite{[Golic]}. He suggested to guess the lower half of all three registers and determine the remaining bits of the registers with the known keystream by Equation \ref{eq}. However, each operation in this attack is much more complicated since it is based on the solutions of system of linear equations. In practice, this algorithm is not better than the Anderson's approach \cite{[And]} or Keller-Seitz's \cite{[KS01]} approach. In deriving the solution of the system of equations, we additionally require solving 44 linear equations by Gaussian Elimination method. This makes Golic's approach impractical to implement.

Pornin and Stern \cite{[Porn]} proposed a \emph{Software-Hardware trade off attack} that is based on Golic's approach. But in contrast to Golic's approach, they guess the clocking sequence at the very beginning. The increased assumptions and complexity of the attacks make the actual implementation very difficult and impractical.

The Biham-Dunkelman attack \cite{[BD]} is expected to be a thousand times faster than the Anderson's attack \cite{[BD]} or Keller-Seitz's attack \cite{[KS01]}. The attack requires $2^{47}$ \emph{A5/1} clockings and about $2^{20.8}$ bits of plaintext data, which is equivalent to 2.36 minutes of conversation. The attacker guesses 12 bits (i.e., $R_{1}[(9,18)]\sim R_{1} [13])$, $R_2[0]$, $R_3[22]$ and $R_3[10]$), and determines the remaining bits of registers $R_1$ and $R_2$ by Equation \ref{eq} and the known keystream bits. The attack algorithm assumes that register $R_3$ is not clocked (i.e., $R_{1}[8] = R_{2}[10] \neq R_{3}[10]$) for 10 consecutive rounds.  Such an event will occur once out of $2^{20}$ possible cipher states. The attacker must know exactly the location of the information-leaking event where register $R_{3}$ is unclocked for 10 consecutive rounds. This is a big assumption. Thus, the attacker will need to probe about $2^{20}$ different starting locations by trial-and-error before the event actually occurs. Also, the probability that such an event, where register $R_{3}$ is not clocked for consecutive 10 rounds occurs is close to zero. This attack requires a lot of data and precomputation space. Hence this attack is not practical for implementation.

Keller and Seitz designed a new attack \cite{[KS01]} based on the attack proposed by Anderson. But unlike Anderson, they took into account the asynchronous clocking of the \emph{A5/1} stream cipher. According to their algorithm, the attacker guesses registers $R_1$ and $R_2$ completely and determines all bits of register $R_3$ by Equation \ref{eq}. The attack was divided into two phases: a \emph{determination phase} in which a possible state candidate consisting of the three registers of \emph{A5/1} after its \emph{warm-up phase} \cite{[Bri]} is generated, and a subsequent \emph{post-processing-phase} in which the state candidate is checked for consistency. In the determination phase, the authors try to reduce the complexity of the simple \emph{guess-and-determine} attack by early recognizing contradictions that could occur by guessing the clocking bit of $R_{3}$ such that $R_{3}$ will not be clocked. Hence, all states arising out of the contradictory guess neither need to be computed further on nor checked afterwards. The authors further reduce the complexity by not only discarding the incorrect possibilities for $R_{3}$[22] in case of contradiction, but also limit the number of choices to the one of not-clocking $R_{3}$, if this is possible without any contradiction. If a case arises where $R_{1}$[8] = $R_{2}$[10] and $R_{3}$[10] has to be guessed, then the authors suggest to always consider the case $R_{1}[8] = R_{2}[10] = R_{3}[10]$ and clock register $R_{3}$ with register $R_{1}$ and register $R_{2}$. This leaves out the possible case of $R_{1}[8] = R_{2}[10] \neq R_{3}[10]$. Thus, the success probability of this attack is approximately 18\%, and the number of state candidates inspected by Keller and Seitz to the number of valid states is $\frac{86}{471} \approx 0.18$.

Gendrullis, Novotny and Rupp \cite{[Rupp]} (GNR) proposed a modification to the Keller-Seitz attack. Unlike Keller-Seitz \cite{[KS01]}, the authors only discard the wrong possibilities for the clocking bit of register $R_{3}$ that would lead to a contradiction. But if no contradiction exists, they check all possibilities of the clocking bit of $R_{3}$, which means the case of clocking and not-clocking $R_{3}$. Thus, every possible state candidate is taken into account, hence giving us a success probability of 100\%.

Besides Golic \cite{[Golic]} and Babbage \cite{[Bab]}, Biryukov-Shamir-Wagner \cite{[Bir]} (BSW) proposed an attack with a complexity of $2^{48}$ requiring about 300GB storage, where the online phase of the attack can be executed within minutes with a $60\%$ success probability. 

Barkan-Biham-Keller \cite{[Bar]} also proposed another attack along these lines.  However, in the precomputation phase of such an attack huge amounts of data need to be computed and stored. For example, with three minutes of ciphertext available, one needs to precompute about 50 TB of data to achieve a success probability of about 60\%. These are practical obstacles that make the implementation of such attacks very difficult.

\section{A New Attack on the \emph{A5/1}}
Our approach is based on the \emph{guess-and-determine attack} proposed by Anderson \cite{[And]}, but with several modifications. With 64 bits of keystream (KS) known, all bits of register $R_{1}$ are guessed (known) and all bits of registers $R_{2}$ and $R_{3}$ are determined. But unlike the approaches of Anderson \cite{[And]}, Golic \cite{[Golic]}, Biham-Dunkelman \cite{[BD]} and Keller-Seitz \cite{[KS01]}, we consider all possible cases i.e., no case is discarded. At the end, we come up with about $2^{48.5}$ possible state candidates, which is much smaller than the exhaustive search where we have $2^{64}$ state candidates. Hence this attack is better than the exhaustive search approach. 
\newline\newline The attack consists of two phases, the \emph{determination phase} and the \emph{post-processing-phase}. The \emph{determination phase} is again divided into two parts, the \emph{processing-phase1} and the \emph{processing-phase2}.

\subsection{Determination Phase} 
\begin{figure}[!htb]
\begin{center}
\hspace*{-1cm}
\includegraphics[trim=0cm 1.6cm 0cm 1cm, clip=true, scale=0.47]{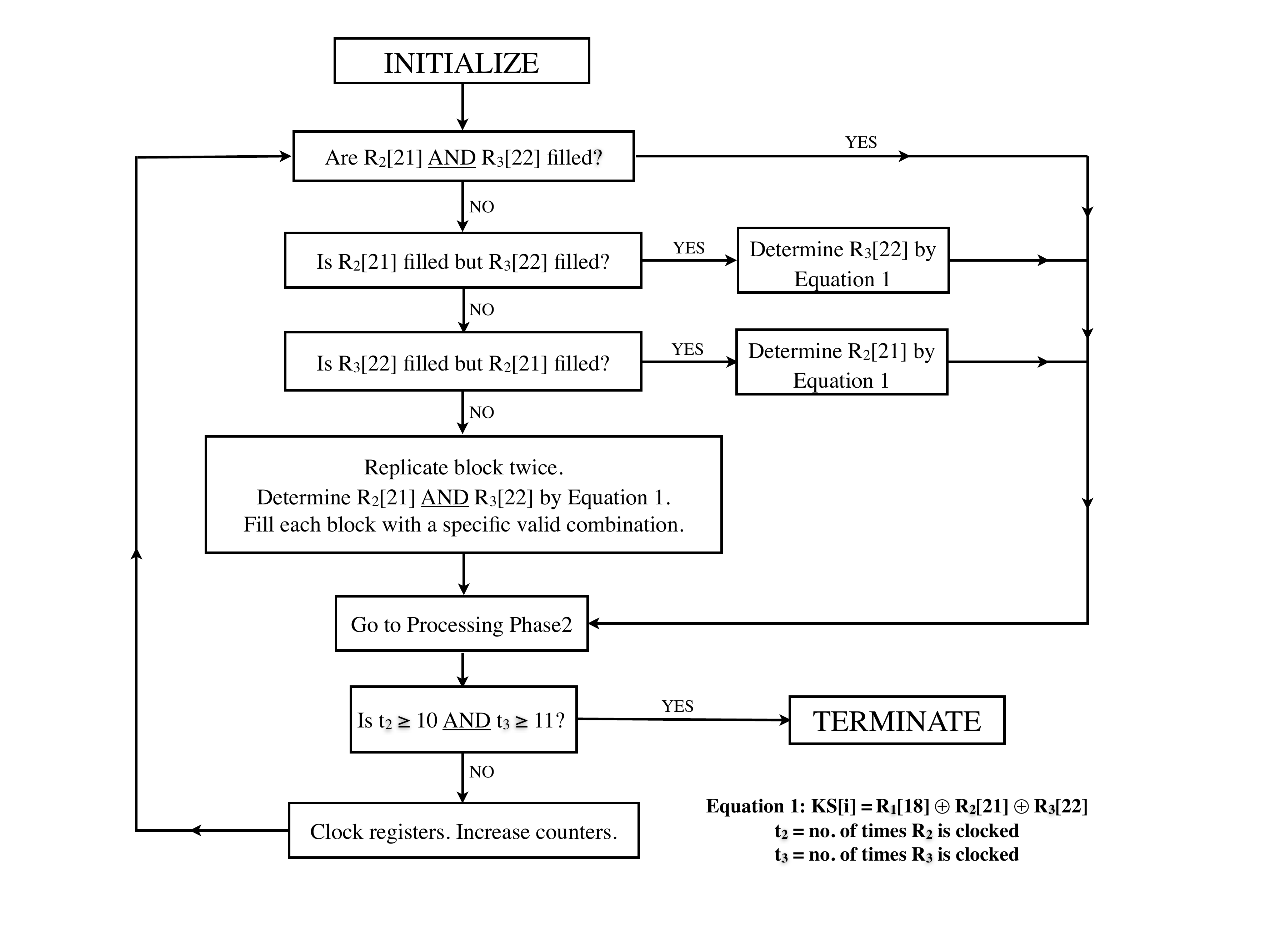}
\caption{Determination Phase of the Attack (Processing-Phase1)}
\label{processing1}
\end{center}
\end{figure}

The \emph{determination phase} generates all possible state candidates after the \emph{warm-up phase} \cite{[Bri]} is completed. 
Let $t_2$ and $t_3$ denote the number of times the registers $R_{2}$ and $R_{3}$ are clocked, respectively. Every time a register is clocked, increment the counter for that register by one. Initialize the algorithm by giving the input of known keystream bits (KS) and guessing all bits of the smallest register $R_{1}$ (Figure~\ref{processing1}).

\subsubsection{Processing-Phase1} Compute the most significant bits (MSBs) of register $R_{2}$ and register $R_{3}$ using the MSB of register $R_{1}$ and KS bit by Equation \ref{eq}.
If the values of three of these bits are known, the fourth can be computed easily by the above equation. If $R_{2}$[21] and $R_{3}$[22] are unknown, then there exist four possible combinations for the unknown bits; i.e. 00, 01, 10 and 11. But the above equation reduced the number of possibilities to two. The two possible combinations that satisfy the equation are: 
\begin{itemize}
\item \emph{If} $R_{1}[18] = KS[i]$, \emph{then} $R_{2}[21] = R_{3}[22] = 0$ or $R_{2}[21] = R_{3}[22] = 1$.
\item \emph{If} $R_{1}[18] \neq KS[i]$, \emph{then} $R_{2}[21] = 0$, $R_{3}[22] = 1$ or $R_{2}[21] = 1$, $R_{3}[22] = 0$.
\end{itemize} 
This reduces the number of possible cases by half and the number of possible state candidates to half. During initialization, $i$ is set to 0, and with every additional clocking round, the value of $i$ increases by 1. Note that $i$ also denotes the total number of clocking rounds that have taken place. 

\subsubsection{Processing-Phase2} Consider the \emph{clocking bits} of registers $R_{2}$ and $R_{3}$. There are three possibilities: 
\begin{itemize}
\item \emph{If} $R_{2}$[10] is filled and $R_{3}$[10] is vacant, \emph{then} replicate the state candidate twice, fill one copy with $R_{3}$[10] = 0, and the other copy with $R_{3}$[10] = 1
\item \emph{If} $R_{2}$[10] is vacant and $R_{3}$[10] is filled, \emph{then} replicate the state candidate twice, fill one copy with $R_{2}$[10] = 0, and the other copy with $R_{2}$[10] = 1
\item \emph{If} $R_{2}$[10] and $R_{3}$[10] are both vacant, \emph{then} replicate the state candidate four times, fill the first copy with $R_{2}$[10] = 0, $R_{3}$[10] = 0; the second copy with $R_{2}$[10] = 0, $R_{3}$[10] = 1; the third copy with $R_{2}$[10] = 1, $R_{3}$[10] = 0; and the fourth copy with $R_{2}$[10] = 1, $R_{3}$[10] = 1.
\end{itemize}

\begin{figure}[!htb]
\begin{center}
\hspace*{-3cm}
\includegraphics[trim=0cm 0.5cm 0cm 0.1cm, clip=true, scale=0.5]{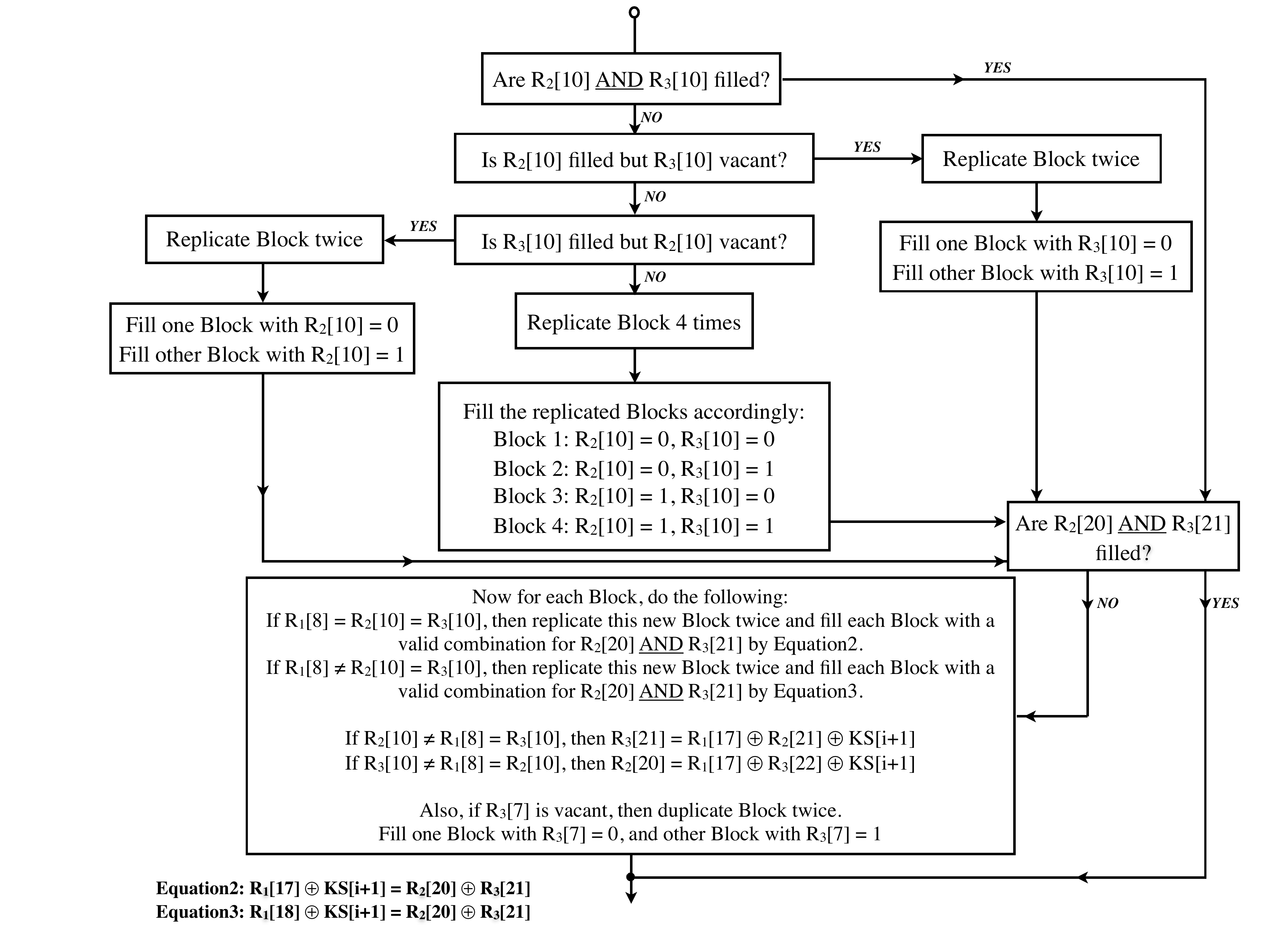}
\caption{Determination Phase of the Attack (Processing-Phase2)}
\label{processing2}
\end{center}
\end{figure}

Thus, all possible combinations are taken into consideration (Figure \ref{processing2}).

Now consider the bits $R_{2}$[20] and $R_{3}$[21]. If registers $R_{2}$ and $R_{3}$ are clocked, then these bits will become the new MSBs for their respective registers after clocking. If both these bits are vacant, there are four possible combinations for these bits; i.e., 00, 01, 10 and 11. But Equation 2 and Equation 3 reduce them to two possibilities. This reduces the number of possible cases by half i.e., 50\% save.

If only one of these bits is vacant, there are two possibilities for the vacant bit i.e., 0 or 1. But this is reduced to only one possibility (Figure \ref{processing2}). For example, if $R_{2}[10] \neq R_{1}[8] = R_{3}[10]$,  then $R_{3}[21] = R_{1}[17] \oplus R_{2}[21] \oplus KS[i+1]$. In this case, only $R_{3}[21]$ is unknown. This bit can be calculated by the above equation. Here, two possibilities for $R_{3}$[21] reduce to only one possibility. This reduces the number of cases by half.

Follow this protocol till $t_2 < 10$ and $t_3 < 11$. Once this condition is not satisfied, i.e., the first time $t_2 \geq 10$ and $t_3 \geq 11$, stop. At this moment, registers $R_{2}$ and $R_{3}$ are completely determined for the known KS and register $R_{1}$. The number of bits between the clocking bit (CB) and the MSB for register $R_{2}$ is 10 and for register $R_{3}$ is 11. Hence, register $R_{2}$ has to be clocked at least 10 times and register $R_{3}$ has to be clocked at least 11 times to determine all the bits of that register.

A \emph{complete state candidate} is a state candidate with all bits filled. The minimum number of KS bits required to obtain a set of complete state candidates is eleven. This will happen when both registers $R_{2}$ and $R_{3}$ are clocked together for 10 consecutive clocking cycles and register $R_{3}$ is clocked again in the next round.

\subsection{Post-Processing-Phase}
The post-processing-phase checks for the key from the set of complete state candidates obtained after the determination phase. As discussed in Section 4.1, the minimum number of rounds needed to perform the post-processing-phase is 11. The number of complete state candidates increases with every additional round. Hence, the probability of finding the key increases with every additional round.

In this phase we generate output bits by performing normal \emph{A5/1} encryption with each of the complete state candidates obtained from the \emph{determination phase}. Match these output bits bit-wise with the known KS bits. If the KS bits and output bits match, continue clocking and generating output bits till a contradiction of bit-wise matching occurs. If all the output bits match the given 64 KS bits, the complete state candidate is the key. Hence, we have found the key among all the complete state candidates.

\section{Analysis of the Attack}
We now discuss each phase of the attack step-by-step. After initialization, we perform the first step of implementation, i.e., the \emph{determination phase}. The state candidate has all bits of register $R_{1}$ and registers $R_2$ and $R_3$ vacant. According to the protocol, the determination phase determines the most significant bits (MSBs) of registers $R_{2}$ and $R_{3}$ in the processing-phase1; the clocking bits of $R_{2}$ and $R_{3}$ (i.e., $R_{2}$[10] and $R_{3}$[10]), bit $R_{3}[7]$ and if possible, bits $R_{2}$[20] and $R_{3}$[21] by processing-phase2. 

Now we analyze the first stage of the determination phase i.e., the processing-phase1. If vacant, the MSBs of registers $R_{2}$ and $R_{3}$ have to be determined. The number of possible combinations reduces from four to two by Equation 1. Thus saving two combinations, i.e., a 50\% save. During the implementation of further rounds, there is a possibility where only one of the MSBs of $R_{2}$ or $R_{3}$ is vacant. We determine these vacant bit(s) by Equation 1.

\begin{figure}[h]
\begin{center}
\includegraphics[trim=1cm 2cm 1cm 0cm, clip=true, scale=0.4]{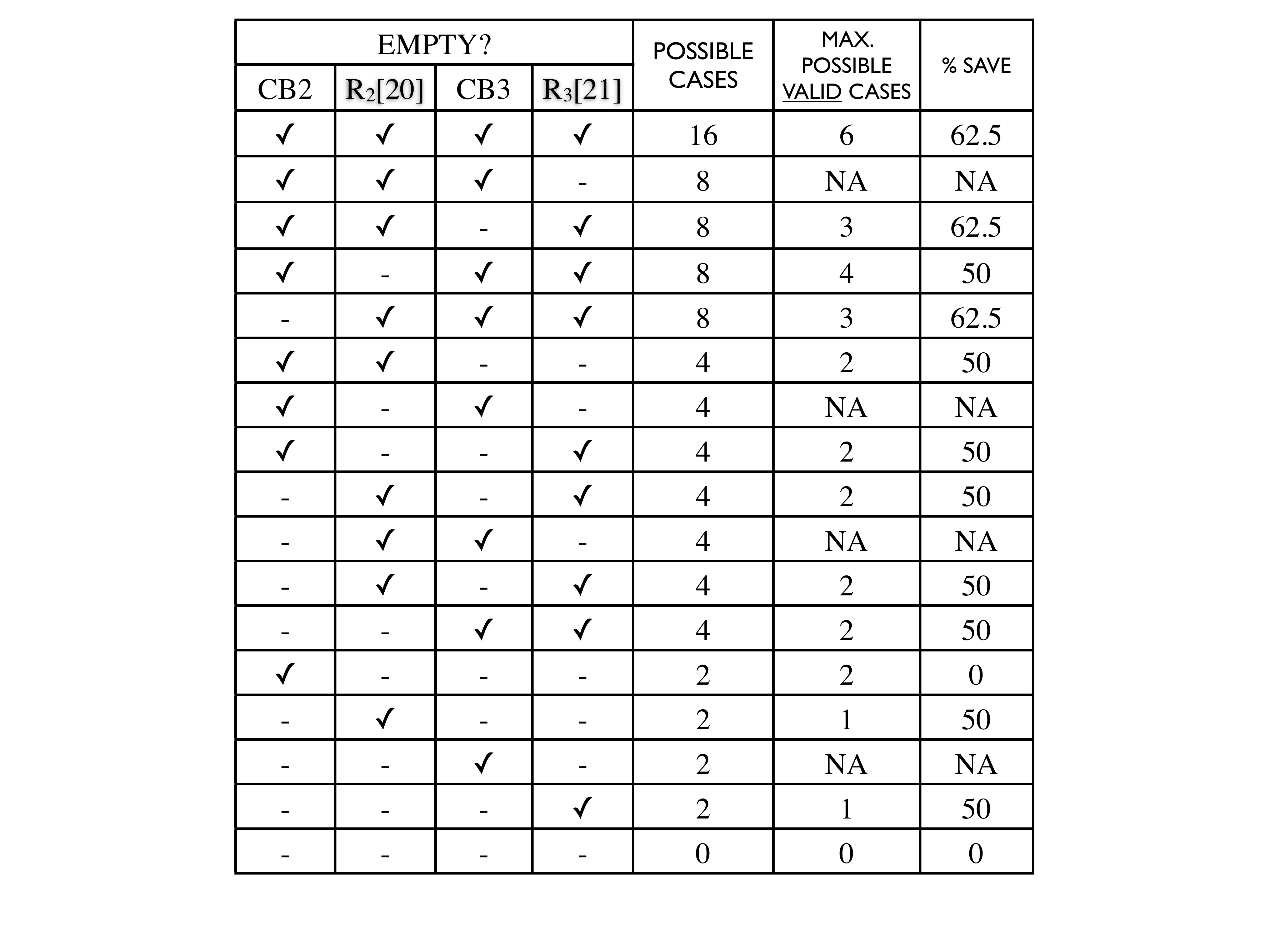}
\caption{All possibilities during Processing-Phase2}
\label{pic}
\end{center}
\end{figure}

We now proceed to processing-phase2 of the determination phase. Here we first consider the four vacant bits: $R_{2}[10]$ (CB of $R_{2}$), $R_{3}[10]$ (CB of $R_{3}$), $R_{2}$[20] and $R_{3}$[21]. But all these four bits (except the first step after initialization) may not be vacant together at all times. In the following table (Figure \ref{pic}), we consider all possible cases of these four bits being empty, and the number of maximum possible valid combinations that exist as a result of Equation 1. We now consider the bit $R_{3}[7]$. There are two possibilities for this bit, i.e., 0 and 1. But we cannot eliminate any case by any method. Hence, we need to consider both cases. 

Whenever the CB of register $R_{3}$ is vacant, the bit $R_{3}[21]$ has to be vacant too. Hence there are some cases in the following table which are \emph{not applicable} (NA). The last column depicts the percentage of the total possible cases that are discarded due to the attack algorithms.

In the determination phase, a total of 7 bits (i.e., $R_{2}[21]$, $R_{2}[20]$, $R_{2}$[10], $R_{3}[22]$, $R_{3}$[21], $R_{3}$[10] and $R_3[7]$) have to be determined. These 7 bits would have $2^7 = 128$ possible combinations. But our algorithms give only 24 valid possible combinations. Thus saving 104 combinations i.e., a saving of 81.25\%. 

If $R_{3}[7]$ is not considered, the first round of implementation will always generate 12 state candidates. On an average, the second round generates 60 state candidates and the third round generates 300 state candidates. The number of state candidates (till round 10) can be approximated by the formula $12 * 5^{n-1}$, where $n$ denotes the $n^{th}$ round, $n \in \mathbb{Z^{+}}$, $n < 11$. It is only after the $11^{th}$ round that we will get the first set of complete state candidates (with all registers full). When bit $R_{3}[7]$ is taken into consideration, the first round of implementation will always generate 24 state candidates. From round three to round ten, the number of possible state candidates after every round is approximately five times the total number in the previous round.

\section{Discussion}
We discuss in detail a probabilistic approach to determine the time complexity, the storage requirement and success probability of our new attack. The results of this probabilistic approach are also corroborated by experimental data.
According to these results, the average number of rounds necessary to get the key is 15.5 and the average number of complete state candidates obtained after 15.5 rounds is $2^{48.5}$.

We conclude this section with a comparison of our attack implementation results with the guess-and-determine attacks already known.

\subsection{Time complexity} 
Now we understand the exact time-complexity of our attack algorithm. According to the algorithm, work done is during:
\begin{itemize}
\item {Checking to see if a cell in a register is vacant or full}
\item {Replication of state candidates}
\item{Filling up vacant bits}
\end{itemize}
It is reasonable to assume that the checking and filling of bits take negligible amount of time. Hence, we can safely assume that the unit of our time complexity measurement should be the number of replications needed, where one unit of time is one replication. The algorithm starts with registers $R_2$ and $R_3$ vacant and register $R_1$ filled (guessed). At the end, it creates about $2^{48.5}$ complete state candidates, i.e., $2^{48.5}$ replications. 

The number of bits between the clocking bit (CB) and the most significant bit (MSB) for register $R_{2}$ is 10 and for register $R_{3}$ is 11. Hence, the number of times the registers $R_{2}$ and $R_{3}$ have to be clocked to determine all the bits of that register is at least 10 and 11 respectively. The minimum number of KS bits required to obtain a set of complete state candidates (with no vacant bits) is 11. This will occur when both registers $R_{2}$ and $R_{3}$ are clocked together for 10 consecutive clocking cycles and register $R_{3}$ is clocked again in the following round. With every clocking round, the number of complete state candidates increases. Hence, the probability of finding the key increases with every additional round of clocking, after 11 rounds.

According to the \emph{majority function} of the \emph{clocking rule} for the \emph{A5/1}, a register will get clocked 3 out of 4 times. At every clocking cycle, at least two registers will get clocked.
Let $n_{1}$ be the event that registers $R_{2}$ and $R_{3}$ are clocked together, and $n_{2}$ be the event that register $R_{1}$ is clocked either with register $R_{2}$ or with register $R_{3}$. The probabilities that events $n_{1}$ and $n_{2}$ occur are given by $P(n_{1})$ and $P(n_{2})$ respectively. Let $n_{2}'$ be the event that only registers $R_{1}$ and $R_{2}$ are clocked. Let $n_{2}''$ be the event that only registers $R_{1}$ and $R_{3}$ are clocked. The probabilities that events $n_{2}'$ and $n_{2}''$ occur are given by $P(n_{2}')$ and $P(n_{2}'')$ respectively. 
Hence, one concludes that $P(n_{1}) = P(n_{2}) = \frac{1}{2}$. Thus:
\begin{equation}
P(n_{1}) + P(n_{2}) = 1 \;\;\text{and}\;\;
P(n_{2}) = P(n_{2}')+ P(n_{2}'') = \frac{1}{4} + \frac{1}{4}.
\end{equation}
i.e.,
\[ P(n_{1}) + P(n_{2}') + P(n_{2}'') = 1 \]
Registers $R_{2}$ and $R_{3}$ have to be clocked at least 10 and 11 times respectively to determine all bits of that register, i.e., to obtain a set of complete state candidates. We assume that they were clocked $n_{1}$ or $n_{2}$ respectively for the attack to stop. 

Let $X$ be the random variable denoting the number of clocking cycles needed to obtain complete state candidate. Let $x_{1}$ be the number of clocking cycles needed for event $n_{1}$, $x_{2}$ for $n_{2}'$ and $x_{3}$ for $n_{2}''$. Here, $x_{1} = 10$, $x_{2} = 10$ and $x_{3} = 11$. Then the expectation for this variable $X$ is defined as
\begin{eqnarray*}
E[X] &= \dfrac{x_{1}* P(n_{1}) + x_{2}* P(n_{2}') + x_{3}* P(n_{2}'')}{P(n_{1}) + P(n_{2}') + P(n_{2}'')}\\ 
&=\dfrac{10* \frac{1}{2} + 2*(10* \frac{1}{4} + 11* \frac{1}{4})}{\frac{1}{2} + \frac{1}{4} + \frac{1}{4}} = 15.5 
\end{eqnarray*}
\subsubsection {An Experiment}
We implement normal encryption of \emph{A5/1} using random inputs for all three registers. The aim of this experiment is to determine the average number of clocking rounds needed for register $R_{2}$ and $R_{3}$ to be clocked at least 10 and 11 times respectively. We performed this experiment thrice, each time with 250 inputs. The average number of clocking rounds needed turned out to be 15.51 with a standard deviation of 1.785. Hence, the experimental results corroborate with the theoretical proof.

So we conclude, that the minimum number of clocking cycles necessary to obtain a set of complete state candidates is 11, and after 15.5 rounds there is a very high probability that the set of complete state candidates contain the key. Experimental results show us that after 11 rounds we get about $2^{40}$ complete state candidates.

\subsection{Storage Requirement and Success Probability} 
As discussed in Section 4.1, the minimum number of KS bits required to generate a set of complete state candidates (all bits filled) is 11. With every additional clocking round, the number of complete state candidates increase. We can start the post-processing-phase of the attack after round 11 simultaneously with the determination phase of the attack. Hence, the probability of finding the key also increases with every round. But we require at least 64 KS bits for the post-processing-phase of the attack to check for the key.

In Table 2, we describe the data obtained from our humble experiments with this attack. The four columns of the table are: number of clocking rounds; total number of state candidates obtained after that round; total number of complete state candidates obtained; and the percentage of complete state candidates over the total number of state candidates for that particular round. All values of the experimental data in the table are approximated to one decimal place.
\begin{table}[!htb]
\caption{Storage Requirement and Success Probability} 
\centering      
\begin{tabular}{c c c c}  
\hline\hline                        
No. of Rounds & Total State Candidates & Complete State Candidates & $\frac{Complete}{Total}\times 100$ \\ [0.5ex] 
\hline                 
11 & $2^{45.2}$ & $2^{39.2}$ & 1.6\% \\   
12 & $2^{46.0}$ & $2^{42.5}$ & 9.0\% \\ 
13 & $2^{46.7}$ & $2^{44.5}$ & 22\%  \\ 
14 & $2^{46.9}$ & $2^{45.3}$ & 30\% \\ 
15 & $2^{47.1}$ & $2^{46.1}$ & 50\% \\ [1ex]       
\hline     
\end{tabular} 
\label{table}  
\end{table}
\paragraph{Remark:}
In each round, the number of possible choices reduce to at least half of the number for an exhaustive search (refer Figure \ref{pic}). Hence, a minimum saving of 50\% takes place in every round over exhaustive search (all possible choices). As stated in Section 6.1, the average number of rounds to get the key is around 15.5. In each round, we save at least half the possible cases. Hence for 15.5 rounds we save $(\frac{1}{2})^{15.5}$ cases. Thus, the average number of complete state candidates obtained after 15.5 rounds will be $(2^{64})(\frac{1}{2})^{15.5} = 2^{48.5}$. The key would be among the set of complete state candidates obtained after the 15th round, with a probability higher than 50\%.
\subsubsection{Storage} 
For each guess of 19 bits of register $R_1$, we require storage of 100MB to determine registers $R_2$ and $R_3$ and the known keystream KS. If we do not find the key for this choice of 19 bits for register $R_1$, then we discard this guess and make a new guess for register $R_1$. In the worst case scenario, we would check all $2^{19}$ possible choices for register $R_1$. The attack has a 100\% success probability and requires about 5.65GB storage.
\subsubsection{Comparison of the known guess-and-determine attacks on the A5/1}
We conclude this section with a comparison of our attack with other guess-and-determine attacks already known. The following table (Table 3) lists the necessary amount of known keystream bits (KS), the time complexity and the success probability for each attack.
\begin{table}[!htb]
\caption{Comparison of the known guess-and-determine attacks on the A5/1} 
\centering      
\begin{tabular}{|c|c| c| c| c|}  
\hline                        
Attack & KS bits & Time  & Success & Notes \\
            &             &complexity &probability & Storage\\[0.5ex]
\hline                 
Anderson's Attack\cite{[And]} & 64 & $2^{52}$ & 100\% & - \\   
Golic's Attack\cite{[Golic]} & 64 & $2^{40.16}$ & 100\% & Additionally solve 64x64 \\ 
 & & & & Linear System of Equations\\
Biham-Dunkelman's Attack\cite{[BD]}& $2^{20.8}$ & $2^{47}$ & 63\% & -  \\ 
Keller-Seitz's Attack\cite{[KS01]} & 64 & $2^{51.24}$ & 18\% & -\\ 
GNR's Attack\cite{[Rupp]} & 64 & $2^{54.02}$ & 100\% & -\\
BSW's Attack\cite{[Bir]} & 64 & $2^{48} $ & 60\% & - \\
Our Attack & 64 & $2^{48.5}$ & 100\% & 5.65GB \\ [1ex]       
\hline     
\end{tabular} 
\label{table}  
\end{table}
\section{Conclusion}
Our attack is based on the \emph{guess-and-determine} approach proposed by Anderson \cite{[And]}, but with several modifications. In this attack, all bits of the first register $R_{1}$ are assumed to be known and all bits of registers $R_{2}$ and $R_{3}$ are determined with 64 bits of the known \emph{keystream} (KS).

This attack has an average time complexity of $2^{48.5}$, which is much smaller than an exhaustive search. The average number of rounds necessary to obtain the correct key from the set of complete state candidates is 15.5. With every round of clocking after 11 rounds, the number of complete state candidates increases. Thus, the probability of finding the key increases with every clocking round. One can do the post-processing phase for a round simultaneously with the determination phase for the next round, and thus save time. The attack is successful with 100\% probability and requires about 5.65GB storage. With this we conclude that our attack is better than all known guess-and-determine attacks on the \emph{A5/1} stream cipher.

\end{document}